# On the Binary Symmetric Channel with a Transition Probability Determined by a Poisson Distribution

A.J. Han Vinck
Univ. Duisburg - Essen, Germany
Univ. of Johannesburg, SA

Fatma Rouissi
Lab. GRES'COM, Ecole
Supérieure des Communications de
Tunis, Tunisia

*Abstract*— The classical Binary Symmetric Channel has a fixed transition probability. We discuss the Binary Symmetric Channel with a variable transition probability that depends on a Poisson distribution. The error rate for this channel is determined and we also give bounds for the channel capacity. We give a motivation for the model based on the Class-A impulse noise model, as given by Middleton. The channel model can be extended to the Additive White Gaussian Channel model, where the noise variance also depends on a Poisson distribution.

## I. INTRODUCTION

The binary symmetric channel (BSC) model [1] is a classical communication model with binary input, binary output and with a fixed transition probability, see Fig. 1.

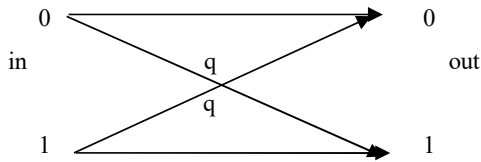

Figure 1. The binary symmetric channel (BSC)

In this case, we calculate the transition probability, q, for a coherent 2-FSK transmission with hard decision (binary) at the receiver [1] using the Q-function, i.e.

$$q = Q\left(\sqrt{\frac{E_b}{2 \times \sigma_g^2}}\right),$$

where $E_b$ is the signal energy per bit and $\sigma_g^2$ is the Additive White Gaussian Noise (AWGN) variance, respectively. $Q(x)$ is the probability that a standard normal random variable takes a value larger than x. We use coherent 2-FSK modulation for the rest of this correspondence.

We modify the model in such a way, that the transition probability is varying according to a Poisson distribution, see Fig. 2.

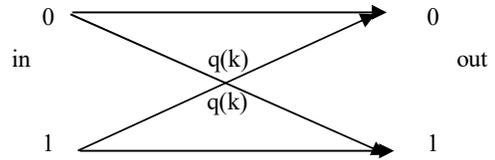

Figure 2. The BSC with variable transition probability

We call this model Channel I. For this model, the transition probability $q(k)$ is given by

$$q(k) = Q\left(\sqrt{\frac{E_b}{2 \times \left(\sigma_g^2 + \frac{k\sigma_I^2}{A}\right)}}\right), \quad (1)$$

where $k$ is Poisson-distributed with the property that the average($k$) = $A$ and the variance ($k$) = $A$. Note: we use the probability $P(k) = (A^k e^{-A}/k!)$ for a Poisson-distributed variable $k$. The noise consists of the sum of two independent AWGN noise sources, both with average zero and sum variance $\sigma_k^2 = \frac{k\sigma_I^2}{A} + \sigma_g^2$. The source with noise variance $\sigma_g^2$ can be considered as a background noise, i.e. always present with the same variance. The variance $\sigma_I^2$ represents the average impulsive noise variance per symbol.

For small values of $A$, a symbol is disturbed with probability $P(k = 0) \approx (1 - A)$ by background noise with variance $\sigma_g^2$, or with probability $P(k = 1) \approx A$ by noise with variance $\sigma_g^2 + \sigma_I^2/A$, which can be large. For large values of $A$, $k$ is close to its average value $A$ and the noise variance approaches $\sigma_g^2 + \sigma_I^2$.

The motivation for this channel is given by the Class-A impulse noise model as defined by Middleton [2]. Using this model, a symbol is disturbed by AWGN in addition to $k$ AWGN impulses, each with variance $\frac{\sigma_I^2}{A}$. Again, $k$ is Poisson distributed with average $A$. At the 2-FSK demodulator, we take a binary decision and end-up with the binary symmetric channel as given in Fig. 2.

We also consider another channel, Channel II, where the transition probability is given by

$$q(k) = Q\left(\sqrt{\frac{E_b}{2\times(\sigma_g^2+k\sigma_I^2)}}\right). \quad (2)$$

Note that here the sum noise variance is $\sigma_g^2 + k\sigma_I^2$, depending on $A$ only via the parameter $k$. For small values of $A$, the noise has the variance $\sigma_g^2$ with $P(k=0) \approx (1-A)$ or the variance $(\sigma_g^2 + \sigma_I^2)$ with $P(k=1) \approx A$. For large values of $A$, $k$ is close to its average value $A$ and thus the variance of the disturbance approaches $\sigma_g^2 + A\sigma_I^2$.

In this correspondence, we investigate the average error probability for Channel I and Channel II for various values of $A$, section II. Based on the calculations, we further look at the channel capacity, section III. We compare the capacity results for informed - and non-informed decoder about the value of k. We conclude in section IV. The Appendix gives an extension of the model to a Gaussian channel with arbitrary input.

## II. ERROR PROBABILITY

To present our results, in the sequel, we set $\sigma_I^2 = 7.28 \cdot 10^{-4}$ Watt, $\sigma_g^2 = 7.28 \cdot 10^{-7}$ Watt and the energy per bit $E_b = 7.28 \cdot 10^{-3}$ Joules [3,4].

The error probability for Channel I (Fig.2) depends on k and can be formulated as

$$BER(channel\ I) = \sum_{k=0}^{\infty} \frac{e^{-A}A^k}{k!} \times Q\left(\sqrt{\frac{E_b}{2\times\left(\sigma_g^2+\frac{k\sigma_I^2}{A}\right)}}\right). \quad (3)$$

Since the Q function is convex for $k \geq 1$, we cannot use the average noise variance $\sigma_g^2 + \sigma_I^2$ to calculate the $BER(channel\ I)$. Instead, we use the properties of the Poisson distribution and first calculate the variance of $(\sigma_g^2 + \frac{k\sigma_I^2}{A})$, which is equal to $(\frac{\sigma_I^4}{A})$ [3,4]. Hence, for large values of $A$ (or small values of $\sigma_I^2$), this variance goes to zero, see Fig. 3.

For large $A$, the relevant contributing probabilities in (3) are close to $P(k = A)$. As result, the $BER(channel\ I)$ approaches:

$$\lim_{A\to\infty} BER(Ch\ I) = Q\left(\sqrt{\frac{E_b}{2\times(\sigma_g^2+\sigma_I^2)}}\right). \quad (4)$$

For small values of $A$, the value of the Q function approaches ½. Hence, in this limiting case, the error rate approaches

$$\lim_{A\to 0} BER(Ch.\ I) = e^{-A}Q\left(\sqrt{\frac{E_b}{2\times(\sigma_g^2)}}\right) + A/2, \quad (5)$$

which is close to the error rate for the AWGN channel with variance $\sigma_g^2$. For small $\sigma_g^2$, the BER is approximately equal to A/2, see Fig. 4. Note that

$$Q\left(\sqrt{\frac{E_b}{2\times(\sigma_g^2)}}\right)(k=0) \leq Q\left(\sqrt{\frac{E_b}{2\times(\sigma_g^2+\sigma_I^2)}}\right).$$

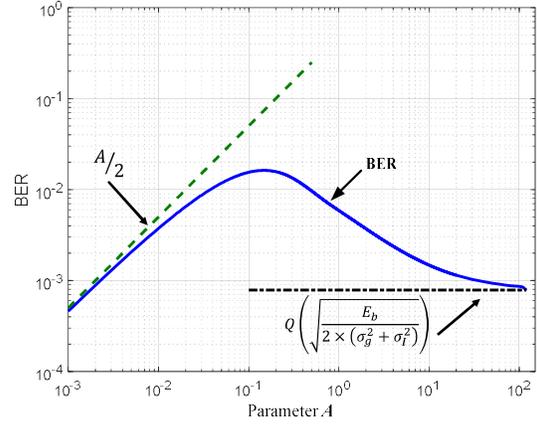

Figure 4. BER versus A, (3)

For our particular choice of $E_b$ and $\sigma_I^2$, an error floor appears for in between values of $A$, see Fig. 5. The SNR is defined as $10\log_{10}(E_b/\sigma_g^2)$ dB, with variable $\sigma_g^2$.

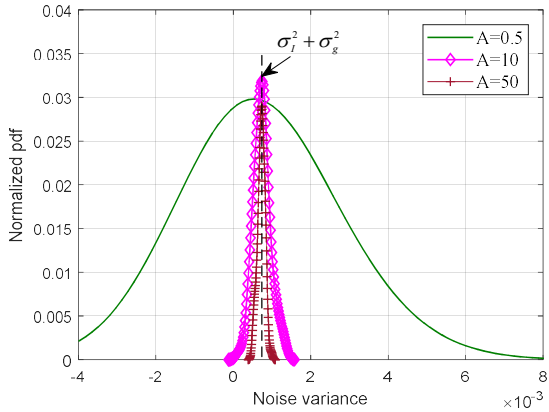

Figure 3. Normalized probability density function of the noise variance for different values of $A$

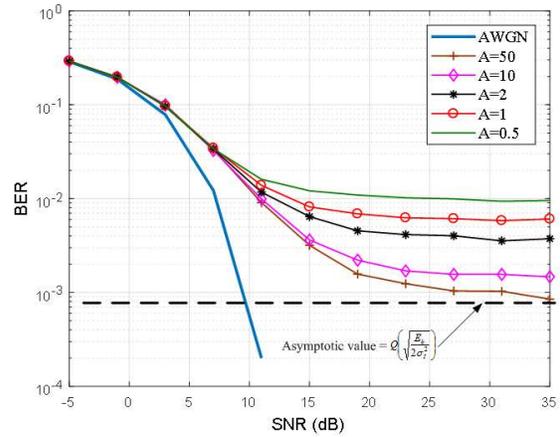

Figure 5. BER versus SNR of a 2-FSK system for different values of $A$

As another varying channel, we can look at the channel II where the noise variance equals $\sigma_g^2 + k\sigma_I^2$, which depends on $k$ and it is again connected to the Poisson distribution with parameter $A$. The BER is given by

$$BER(Ch.\ II) = \sum_{k=0}^{\infty} \frac{e^{-A}A^k}{k!} \times Q\left(\sqrt{\frac{E_b}{2\times(\sigma_g^2 + k\sigma_I^2)}}\right). \quad (6)$$

For large values of $A$, the relevant values of $k$ are large and thus, we expect $Q\left(\sqrt{\frac{E_b}{2\times(\sigma_g^2+k\sigma_I^2)}}\right) \approx \frac{1}{2}$ and

$$\lim_{A\to\infty} BER(Ch.II) = 1/2. \quad (7)$$

For small values of $A$, $P(k=0) = e^{-A} \approx 1 - A$, and thus we expect the BER to be close to

$$\lim_{A\to 0} BER(Ch.II) =$$

$$(1-A)Q\left(\sqrt{\frac{E_b}{2\times(\sigma_g^2)}}\right) + A\,Q\left(\sqrt{\frac{E_b}{2\times(\sigma_g^2+\sigma_I^2)}}\right). \quad (8)$$

The value of $\sigma_g^2$ is small compared to $\sigma_I^2$, and thus for small $A$, the $BER(Ch.II, small\ A)$ is approximately equal to $AQ\left(\sqrt{\frac{E_b}{2\times(\sigma_g^2+\sigma_I^2)}}\right)$. In Fig. 6, we illustrate the BER for channel II.

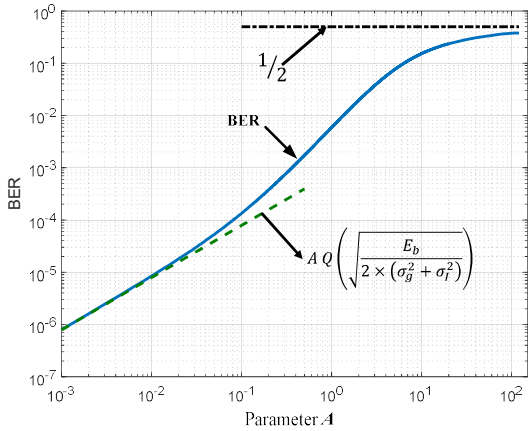

Figure 6.   The BER for channel II as a function of $A$.

The BER can be used to estimate the channel capacities for Channel I and Channel II, section III.

### III. CAPACITY

In Information theory, the channel capacity is an important part of channel evaluation, [5,6]. For the binary symmetric channel, the capacity is $C = 1 - H(q)$, where $H(q)$ is the binary entropy function, with a maximum value of 1. Channel I and Channel II are symmetric and thus the capacity achieving input probability distribution $P(in) = (½, ½)$. In general, the capacity is given by $H(Y) - H(Y/X)$, where $X$ and $Y$ are the input- and output – random variables, respectively. For these two channels and $P(in) = (½, ½)$, the output probability distribution is $P(out) = (½, ½)$. Hence, to find the capacity, we have to find the entropy $H(Y/X) = H(Noise)$. When receiver and transmitter are <u>not</u> informed about the value of $k$, the $H(Noise) = H(BER)$. The channel capacity is then $C = 1 - H(BER)$.

Suppose that the receiver is informed about the value of $k$ (state of the channel), then the capacity is given by

$C_{informed}$ = $P(k = 0)C(k = 0) + P(k = 1)C(k =1 ) + ….$
 = $1 - P(k = 0)H(q(k = 0)) - P(k = 1)H(q(k = 1)) + …$
 = $1 - \sum_{k=0}^{\infty} P(k)\,H(q(k))$ (9)
 $\geq 1 - H(\overline{q(k)})$ (using the entropy convexity) (10)

where $q(k) = Q\left(\sqrt{\frac{E_b}{2\times\left(\sigma_g^2 + \frac{k\sigma_I^2}{A}\right)}}\right)$ and $\overline{q(k)} = BER$ as given in (3) for channel I and (6) in channel II.

Fig. 7 illustrates the capacities for Channel I for different values of A for informed and non-informed receiver.

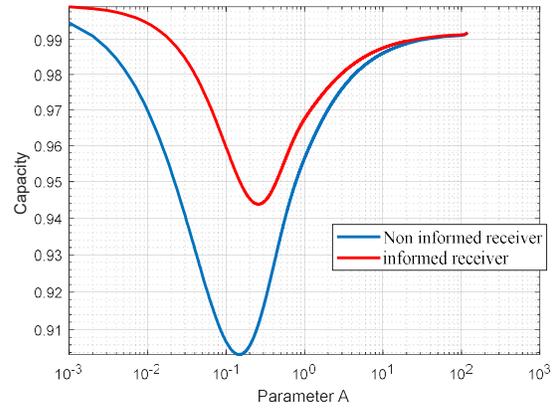

Figure 7.   Capacity for Channel I as a function of $A$.

For large values of $A$, the capacity is determined by (4), since the variance of $\left(\frac{k\sigma_I^2}{A}\right) = \left(\frac{\sigma_I^4}{A}\right)$, which goes to 0 for large $A$. Hence, information at the receiver about $k$ does not help since all relevant values of $k$ are close to the average $A$ and thus the BER is equal to (4).

For small values of $A$, the information at the receiver does help. The value of $\sum_{k=0}^{1} P(k)\,H(q(k)) \approx A$, whereas $H(\overline{q(k)}) \approx H(A/2)$. Note, that $(1-A) > 1 - H(A/2)$ for $A < 1$. As an example, we calculate (9) and (10) when $A = 0.01$. In this case, for the given noise variances, we obtain $1 - \sum_{k=0}^{\infty} P(k)\,H(q(k)) \approx 0.99$ and $1 - H(\overline{q(k)}) \approx 0.96$.

Remark: From Fig. 7, we can see that for the non-informed receiver the capacity for $A = 0.02$ is the same as for the informed receiver at $A = 0.2$. Hence, in order to obtain the same performance, we have to increase $E_b$ for the non-informed model with a factor of 10. This shows the importance of channel information at the receiver.  Noise mitigation is an important topic. In [7], we find an approach using noise estimation at the receiver.

Fig. 8 illustrates the capacities for Channel II for different values of A for informed and non-informed receiver. The values of (9) and (10) are very close together. The gain by using information at the receiver is negligible. For small $A$, (8) approaches $AQ\left(\sqrt{\frac{E_b}{2\times(\sigma_g^2+\sigma_I^2)}}\right)$. For large $A$, the variance gets small and thus, all values of $k$ are close to the average $A$ leading to a BER approaching ½ and a capacity close to 0.

Note: This channel capacity analysis concerns the particular case of a discrete symmetric channel, on which the capacity expression is a function of *BER*, and cannot exceed a value of 1. Further details of the channel capacity for a general AWGN channel are provided in the Appendix.

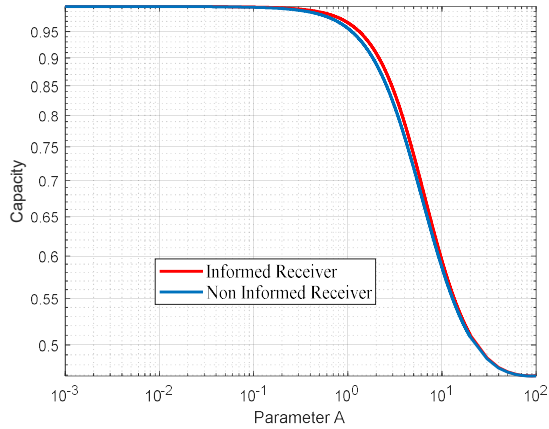

Figure 8. Capacity for Channel II as a function of *A*.

## V CONCLUSION

We introduce a variation of the Binary Symmetric Channel (BSC), where the transition probability is determined by a Poisson distribution with characteristic parameter *A*. Two different types of channels are described. We give the Bit Error Rate (BER) and the channel capacity for informed and non-informed receiver about the strength of the noise that determines the transition probability. Using the *BER*, we also comment on the channel capacities for both cases. We extend the model to the AWGN channel.

ACKNOWLEDGMENT

We thank Prof. Adel Ghazel for his continuous support.

## APPENDIX

The Binary Symmetric channel can be generalized to the un-quantized Additive White Gaussian Noise channel with noise variance $(\sigma_g^2 + \frac{k\sigma_I^2}{A})$, see Fig. 9.

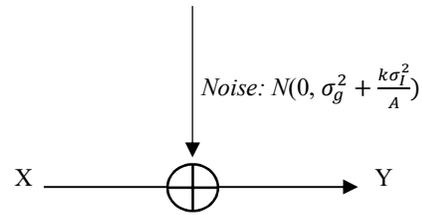

Figure 9. General AWGN channel.

The average variance is $\sigma_{AV} = \sigma_g^2 + \sigma_I^2$ and the average input power spectral density is *P/2B*, where *P* and *B* are the transmitted signal power and bandwidth, respectively.

Calculating the channel capacity means that for a given noise probability distribution *p(n)* the input distribution *p(x)* has to be found which maximizes

$$C = \sup_{p(x):\ \overline{x^2}\ \leq\ P/2B} [H(Y) - H(N)] \text{ bits/transmission}, \quad (10)$$

where $H(Y)$ and $H(N)$ are the output- and - noise entropy, respectively. Note that $H(N) = H(Y/X)$. For an AWGN source with fixed variance $\sigma_g^2$, $H(N)= ½log_2(\sigma_G^2)\ bits$. Unfortunately, it is not always possible to find an explicit expression for the optimum input distribution that maximizes *C*, [6].

An interesting question is whether information about the strength of the noise, $\frac{k\sigma_I^2}{A}$, at the receiver and/or the transmitter has an influence on the capacity of the channel. This approach has led to interesting results for the capacities of memory systems, [8]. We look at the four possible different situations of informed or non-informed transmitter and –receiver, respectively (where * can be + or -):

1. The capacity C(+,*): the transmitter knows the value of *k*. In this case, the transmitter can use the water-filling principle which maximizes the output entropy to $H(N) = H(P/2B + \sigma_{AV}^2)$ bits;

2. The capacity C(*,+): The <u>receiver knows</u> the value of *k*. The entropy of the noise at the output is given by

$$H(N) = \sum_{k=0}^{\infty} P(k) H(\sigma_g^2 + \frac{k\sigma_I^2}{A}) \leq H(\sigma_{AV}^2) \text{ bits};$$

3. The capacity C(*,-): the receiver does not know the value of *k* and thus,

$$H(N) = H[\sum_{k=0}^{\infty} P(k)(\sigma_g^2 + \frac{k\sigma_I^2}{A})] = H(\sigma_{AV}^2) \text{ bits};$$

4. The capacity C(-,*): The transmitter does not know the value of *k* and always uses the maximum input power spectral density P/2B. The output entropy is given by

$$H(Y) = \sum_{k=0}^{\infty} P(k) H(P/2B + \sigma_g^2 + \frac{k\sigma_I^2}{A}) \text{ bits}.$$

In conclusion, the respective capacities C(+,+), C(-,+), C(-,-), and C(+,-) are given by:

$$C(+,+) = H(P/2B + \sigma_{AV}^2) +$$
$$- \sum_{k=0}^{\infty} P(k)[H(\sigma_g^2 + \frac{k\sigma_I^2}{A})] \text{ b/tr} ; \quad (11)$$

$$C(+,-) = H(P/2B + \sigma_{AV}^2) - H(\sigma_{AV}^2) \text{ b/tr} ; \quad (12)$$

$$C(-,+) = \sum_{k=0}^{\infty} P(k)[H(P/2B + \sigma_g^2 + \frac{k\sigma_I^2}{A}) +$$
$$- H(\sigma_g^2 + k\frac{\sigma_I^2}{A})] \text{ b/tr} ; \quad (13)$$

$$C(-,-) = \sum_{k=0}^{\infty} P(k)[H(P/2B + \sigma_g^2 + \frac{k\sigma_I^2}{A}) +$$
$$- H(\sigma_{AV}^2) \text{ b/tr}. \quad (14)$$

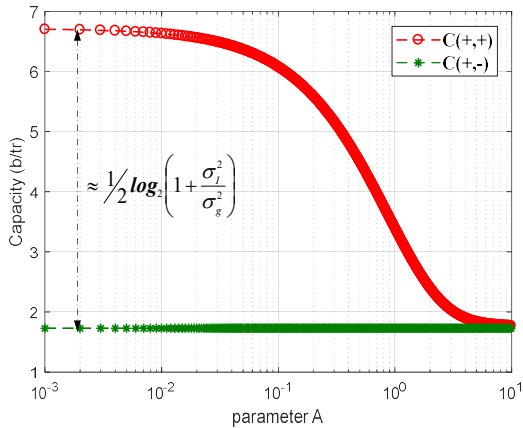

Figure 10. C(+,+) & C(+,-) versus parameter *A*.

For <u>large values of *A*</u>, the variance of $\frac{k\sigma_I^2}{A}$ goes to zero, *k* will be close to *A* and thus C(+,+), C(-,+), C(-,-), and C(+,-) approach, (Figs. 10 and 11).

$$\lim_{A \to 0} C(*,*) = H(P/2B + \sigma_{AV}^2) - H(\sigma_{AV}^2) \text{ b/tr}. \quad (15)$$

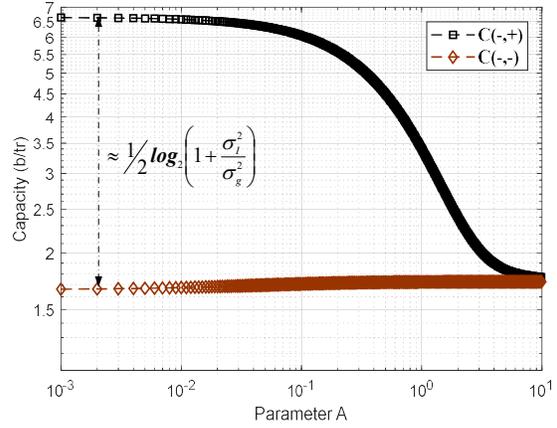

Figure 11. C(-,+) & C(-,-) versus parameter *A*.

For <u>small values of A</u>, $P(k = 0)$ and $P(k = 1)$ can be approximated by *(1-A)* and *A*, respectively. Using the definition for the entropy, the factor ½ $log_2(1 + \frac{\sigma_I^2}{\sigma_g^2})$ dominates the difference between informed and non-informed receiver capacity, i.e.

$$\lim_{A \to 0} [C(*,+) - C(*,-)] =$$
$$= H(\sigma_{AV}^2) - \sum_{k=0}^{\infty} P(k)[H(\sigma_g^2 + \frac{k\sigma_I^2}{A})] \text{ b/tr}.$$

The difference between C(+,*) and C(-,*) =

$$\lim_{A \to 0} [C(+,*) - C(-,*)] = H(P/2B + \sigma_{AV}^2) +$$
$$- \sum_{k=0}^{\infty} P(k)[H(P/2B + \sigma_g^2 + \frac{k\sigma_I^2}{A}) \text{ b/tr}.$$

For small values of A, the difference [C(+,*) - C(-,*)] is dominated by the factor ½ $log_2(1 + \frac{\sigma_I^2}{(\frac{P}{2B}) + \sigma_g^2})$. See Fig. 12.

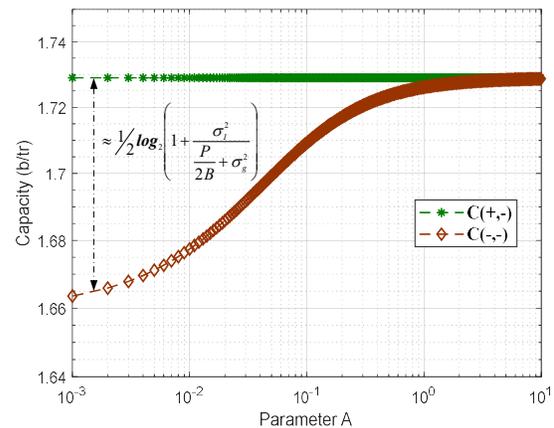

Figure 12. C(+,-) & C(-,-) versus parameter *A*.

We conclude that for small *A*, the difference in capacity is larger for the informed receiver than for the informed transmitter. For large *A*, the four capacities merge.